\def \SAIT #1 #2 {{\em Mem.\ Soc.\ Astron.\ It.\/} {\bf #1}, #2}
\def \MESS #1 #2 {{\em The Messenger\/} {\bf #1}, #2}
\def \ASTRNACH #1 #2 {{\em Astron. Nach.\/} {\bf #1}, #2}
\def \AAP #1 #2 {{\em Astron. Astrophys.\/} {\bf #1}, #2}
\def \AAL #1 #2 {{\em Astron. Astrophys. Lett.\/} {\bf #1}, L#2}
\def \AAR #1 #2 {{\em Astron. Astrophys. Rev.\/} {\bf #1}, #2}
\def \AAS #1 #2 {{\em Astron. Astrophys. Suppl. Ser.\/} {\bf #1}, #2}
\def \AJ #1 #2 {{\em Astron. J.\/} {\bf #1}, #2}
\def \ANNREV #1 #2 {{\em Ann. Rev. Astron. Astrophys.\/} {\bf #1}, #2}
\def \APJ #1 #2 {{\em Astrophys. J.\/} {\bf #1}, #2}
\def \APJL #1 #2 {{\em Astrophys. J. Lett.\/} {\bf #1}, L#2}
\def \APJS #1 #2 {{\em Astrophys. J. Suppl.\/} {\bf #1}, #2}
\def \APSS #1 #2 {{\em Astrophys. Space Sci.\/} {\bf #1}, #2}
\def \ASR #1 #2 {{\em Adv. Space Res.\/} {\bf #1}, #2}
\def \BAIC #1 #2 {{\em Bull. Astron. Inst. Czechosl.\/} {\bf #1}, #2}
\def \JSQRT #1 #2 {{\em J. Quant. Spectrosc. Radiat. Transfer\/} {\bf #1}, #2}
\def \MN #1 #2 {{\em Mon. Not. R. Astr. Soc.\/} {\bf #1}, #2}
\def \MEM #1 #2 {{\em Mem. R. Astr. Soc.\/} {\bf #1}, #2}
\def \PRL #1 #2 {{\em Phys. Rev. Lett.\/} {\bf #1}, #2}
\def \PLR #1 #2 {{\em Phys. Lett. Rev.\/} {\bf #1}, #2}
\def \PASJ #1 #2 {{\em Publ. Astron. Soc. Japan\/} {\bf #1}, #2}
\def \PASP #1 #2 {{\em Publ. Astr. Soc. Pacific\/} {\bf #1}, #2}
\def \NAT #1 #2 {{\em Nature\/} {\bf #1}, #2}

\documentstyle{memsait}
\input epsf.sty
\input psfig.sty
\begin{opening}
\title{AGILE:
 A GAMMA-RAY MISSION FOR A LIGHT IMAGING DETECTOR$^{(\star)}$}

\vspace*{-.5cm}

\author{M.TAVANI$^{1,2}$, G.BARBIELLINI$^3$, P.CARAVEO$^1$, 
S.DI PIPPO$^4$,
S.MEREGHETTI$^1$,
A.MORSELLI$^5$, A.PELLIZZONI$^1$, A.PERRINO$^5$, P.PICOZZA$^5$,
P.SCHIAVON$^3$, S.SEVERONI$^5$, F.TAVECCHIO$^1$, A.VACCHI$^3$,
S.VERCELLONE$^1$}
\institute{$^1$Istituto di Fisica Cosmica "G.Occhialini", CNR, Milano, Italy\\
$^2$Astrophysics Laboratory, Columbia University, New York, USA \\
$^3$Dipartimento di Fisica, Universit\`a di Trieste e INFN, Italy \\
$^4$Agenzia Spaziale Italiana \\
$^5$Dipartimento di Fisica,Universit\`a di Roma II, 
"Tor Vergata" e  INFN, Italy}
\date{} 
\end{opening}

\begin{document}

\oddpagefooter{}{}{} 
\evenpagefooter{}{}{} 
\ 
\bigskip

\vspace*{-1.3cm}

\begin{abstract}
AGILE is an innovative, cost-effective gamma-ray mission
proposed to the ASI Program of Small Scientific Missions.
It is planned to  detect  gamma-rays in the 30~MeV--50~GeV energy band
and operate as an {\bf Observatory} open to the international
community. Primary scientific goals  include the study of 
AGNs, gamma-ray bursts, Galactic sources, unidentified gamma-ray sources,
solar flares, and diffuse gamma-ray emission.
 AGILE is planned to be operational during
the year 2001 for a 3-year mission. It  will ideally 
 `fill the gap' between  EGRET  and GLAST, 
and  support ground-based
  multiwavelength studies of high-energy sources.
\end{abstract}

\newcommand{\nn}{\noindent}
\newcommand\etal{{et al.}}
\newcommand{\ci}[1]{\cite{#1}}
\newcommand{\cit}[1]{\cite{#1}}
\newcommand{\loe}{\stackrel{<}{\sim}}
\newcommand{\goe}{\stackrel{>}{\sim}}
\def\agile{ AGILE }

%
\def\jref#1 #2 #3 #4 {{\par\noindent \hangindent=3em \hangafter=1
      \advance \rightskip by 0em #1, {\it#2}, {\bf#3}, #4.\par}}
\def\rref#1{\begin{flushleft}{\par\noindent \hangindent=-3em \hangafter=1
      \advance \rightskip by 0em #1.\par}\end{flushleft}}


%
%
%
 
\section{Introduction }

 The gamma-ray mission AGILE ({\it Astro-rivelatore Gamma a Immagini LEggero})
is currently in a study phase for the Italian Space Agency (ASI)
program of Small Scientific Missions.
AGILE ideally conforms to the {\it faster, cheaper,  better} philosophy
adopted by space agencies  for scientific missions.
Gamma-ray detection by AGILE 
  is  based on  silicon tracking  detectors developed
for space missions  by INFN and Italian University laboratories 
during the past years (Barbiellini et al., 1995; Morselli et al., 1995).
 AGILE is both very light ($\sim 60$~kg) and  highly  efficient in
detecting and monitoring gamma-ray sources 
in the energy range 30~MeV--50~GeV. The accessible field of
view is unprecedently large ($\goe  1/5$ of the whole sky)
because of state-of-the-art readout electronics and segmented
anticoincidence system.
AGILE was selected by ASI (1997 December)
for a  phase A study to be completed within the end of 1998.
The goal is to achieve an on-axis sensitivity comparable to that of 
EGRET on board of CGRO
(a smaller background resulting from an improved  angular resolution 
 more than compensates the loss due to  a smaller effective area)
and a better sensitivity for large off-axis angles (up to $\sim 60^{\circ}$).
Planned to be operational during the year 2001 for a 3-year mission,
AGILE will ideally `fill the vacuum' between the end of EGRET operations and
 GLAST.
AGILE's data will provide 
crucial support for ground-based observations and
 several space missions including
AXAF, INTEGRAL, XMM, ASTRO-E, SPECTRUM-X.

\noindent
------------------------------------------------------------------------------- \\
($\star$) Adapted from a paper presented at the Conference
{\it Dal nano- al Tera-eV: tutti i colori degli AGN}, Rome 18-21 May 1998,
to be published by the {\sl Memorie della Societa' Astronomica Italiana}.
\newpage

Fig.~1 (left panel) shows the baseline instrument layout.
We refer to a companion  paper for more details on
the instrument (Morselli et al., 1998).
Spectral information ($\Delta \, E/E \sim 1$)
will be obtained by multiple scattering
of created pairs in tungsten-silicon planes (for energies less than
$\sim 500$~MeV) and by the use of a  mini-calorimeter.
We are also studying the possibility of adding 
 an ultra-light coded mask imaging system sensitive in the energy band
$\sim $10--40 keV  on top of AGILE.
Super-AGILE is an innovative concept\footnote{Developed in 
collaboration with E. Costa, M. Feroci, L. Piro and P. Soffitta},
 combining silicon technology
to simultaneously detect gamma-rays and hard X-rays with accurate imaging.

\section{Scientific Objectives}

Table~1 summarizes the expected performance of AGILE vs. that of
EGRET.
Because of the large field of view
 ($\sim 0.8 \,\pi$~sr)
AGILE will discover a large number
of gamma-ray transients, monitor known sources, and allow rapid
multiwavelength follow-up observations because of a dedicated
data analysis and alert program. Fig.~1 (right panel) shows the off-axis
response to gamma-ray detection of AGILE and EGRET.
We summarize here AGILE's scientific objectives.

$\bullet$ {\it Active Galactic Nuclei}.
For the first time,
simultaneous monitoring of a large number of AGNs per pointing
will be possible.
Several outstanding issues concerning the mechanism
of AGN gamma-ray production  and activity can be addressed by AGILE
 including:
(1) the study of transient vs. low-level gamma-ray emission and duty-cycles;
(2) the relationship between the gamma-ray variability and
the radio-optical-X-ray-TeV emission;
(3) the correlation between relativistic radio plasmoid ejections
and gamma-ray flares.
A  program for joint AGILE and  ground-based monitoring observations
is being planned.
On the average, AGILE will achieve  deep  exposures of AGNs
and  substantially improve our knowledge on the low-level emission 
 as well as detecting flares.
We conservatively estimate that for a 3-year program
AGILE will detect a number of AGNs 2--3 times larger than that of EGRET.
A companion paper presents 
the impact of AGILE on the study of AGNs (Mereghetti et al., 1998).
Super-AGILE will monitor, for the first time,  simultaneous
 AGN emission in the gamma-ray and
hard X-ray ranges.

$\bullet$ {\it Diffuse Galactic  and  extragalactic emission}.
The AGILE good angular resolution
and large average exposure  will further
improve our knowledge of cosmic ray origin, propagation,
interaction and emission processes. We also note that a
joint study of gamma-ray emission from MeV to TeV energies
is possible by special  programs involving AGILE and
new-generation TeV observatories of improved angular resolution.

$\bullet$ {\it Gamma-ray pulsars}.
\agile  will contribute to the study of gamma-ray pulsars
in several ways:
(1) improving photon statistics for gamma-ray period searches
 by  dedicated observing programs
with long observation  times of 1-2 months per source;
(2) detecting possible secular fluctuations of the gamma-ray
emission from neutron star magnetospheres;
(3) studying unpulsed gamma-ray emission from plerions in
supernova remnants and searching for time variability of
pulsar wind/nebula interactions, e.g., as in  the  
Crab nebula (de Jager et al., 1996).

$\bullet$ {\it Galactic sources, new transients}.
A large number of gamma-ray sources near  the
Galactic plane are unidentified, and sources such as
2CG~135+1 or transients (e.g., GRO~J1838-04) can be
monitored on timescales of months/years.
Also Galactic X-ray jet sources (such as Cyg X-3,
GRS~1915+10, GRO~J1655-40 and others) can produce detectable
gamma-ray emission for favorable jet geometries,
and a TOO program is planned to follow-up new
discoveries of {\it micro-quasars}.

$\bullet$ {\it Gamma-ray bursts}. 
 About ten GRBs have been
detected by EGRET's spark chamber during $\sim 7$
years of operations (Schneid et al., 1996a).  This number 
appears to be  limited
by the EGRET FOV and sensitivity and  not by the
GRB emission mechanism.
GRB detection rate by AGILE is expected to be 
a factor of $\sim 5$ larger than that of EGRET,
i.e.,  $\goe$5--10 events/year).
 The small AGILE deadtime ($\goe 100 $ times smaller than
that of EGRET) allows a  better study of the
initial phase of GRB pulses (for which EGRET response was
in many cases inadequate). The remarkable 
discovery  by EGRET of `delayed'  gamma-ray emission up to  $\sim 20$~GeV
from GRB~940217 (Hurley et al., 1994)
is of great  importance to model burst  acceleration processes.
 AGILE is expected to be  highly efficient
in detecting photons above 10~GeV because of limited backscattering.
Super-AGILE will be able to locate GRBs within a few arcminutes,
and will systematically study the interplay between hard X-ray
 and gamma-ray emissions.

$\bullet$ {\it Solar flares}.
During the last solar maximum,
solar flares were  discovered to produce prolonged high-intensity
gamma-ray outbursts 
(e.g., Schneid et al., 1996b).
AGILE will be operational during part of the next solar maximum
and several solar flares may be detected. Particularly important
for analysis 
will  be the flares simultaneously detected by AGILE and HESSI 
(sensitive in the band 20~keV--20~MeV).

\begin{table}          
\vspace*{-0.3cm}
{
\begin{center}
 \centerline{\bf Table 1:  Comparison between AGILE and EGRET}
\vskip .03in \begin{tabular}{l l l}\hline    & EGRET & AGILE \\ \hline
Mass (kg) & {1830 } & {60 } \\ \hline
Energy band & 30 MeV -- 30 GeV & 30 MeV -- 50 GeV \\ \hline
Field of view (sr)  & $0.15~\pi $ & $ 0.8~\pi $ \\ \hline 
Angular resolution$^{*}$   & $\loe 1^{\circ}$ & $\loe 0.5^{\circ}$ \\ \hline Sensitivity  & 
${8 \times 10^{-9}}$ & 6$\times$ 10$^{-9}$ 
(@ $0.1$ GeV)\\ for pointlike sources$^{\dag}$ & 
${1 \times 10^{-10}}$ & 4$\times$ 10$^{-11}$ (@ $1$ GeV)\\
($\rm ph \, cm^{-2} \, s^{-1} \, MeV^{-1}$) &${1 \times 10^{-11}}$  
& 3 $\times$10$^{-12}$ 
(@ $10$ GeV) \\ \hline 
Required pointing reconstruction & $\sim$10 arcmin & $\sim$1-2 arcmin\\  \hline \end{tabular}
\end{center}

{\small 
(*)
 FWHM of the point spread function ($E > 100$~MeV)  calculated for
an incidence angle less than $20^{\circ}$ and
  a photon spectrum  $\sim E^{-2}$).
($\dag$) Obtained for a typical exposure time near 2 weeks at high galactic
latitude for both AGILE and EGRET.}
}
\end{table}

\section{Mission}

AGILE is planned to be integrated  with a spacecraft
of the MITA class currently being developed by  Gavazzi Space
with the support of ASI. AGILE's pointing is obtained
by a three-axis stabilization system with an
accuracy near 0.5--1 degree. Pointing reconstruction
reaching an accuracy of 1--2 arcmin is obtained by star trackers.
The downlink telemetry rate is planned to be $\sim 500 \; \rm kbit \, s^{-1}$,
and  is adequate for AGILE and Super-AGILE
 for a single contact per orbit. The ideal orbit is
equatorial (550 -- 650~km).

The AGILE mission is being planned as an {\bf Observatory}
open to the international scientific community.
Planning of pointed observations, quicklook and standard
data analysis results will be  available  to
the community through a Guest Observer
Program. The AGILE mission emphasizes a rapid
response to the detection of gamma-ray transients.
The  AGILE Science Support Group will help coordinating
multiwavelength observations of gamma-ray
sources, and will stimulate investigations
of observational and theoretical nature on gamma-ray sources
detected by AGILE.

\begin{figure}
\setbox111=\hbox{\hspace*{-2.cm}
\psfig{figure=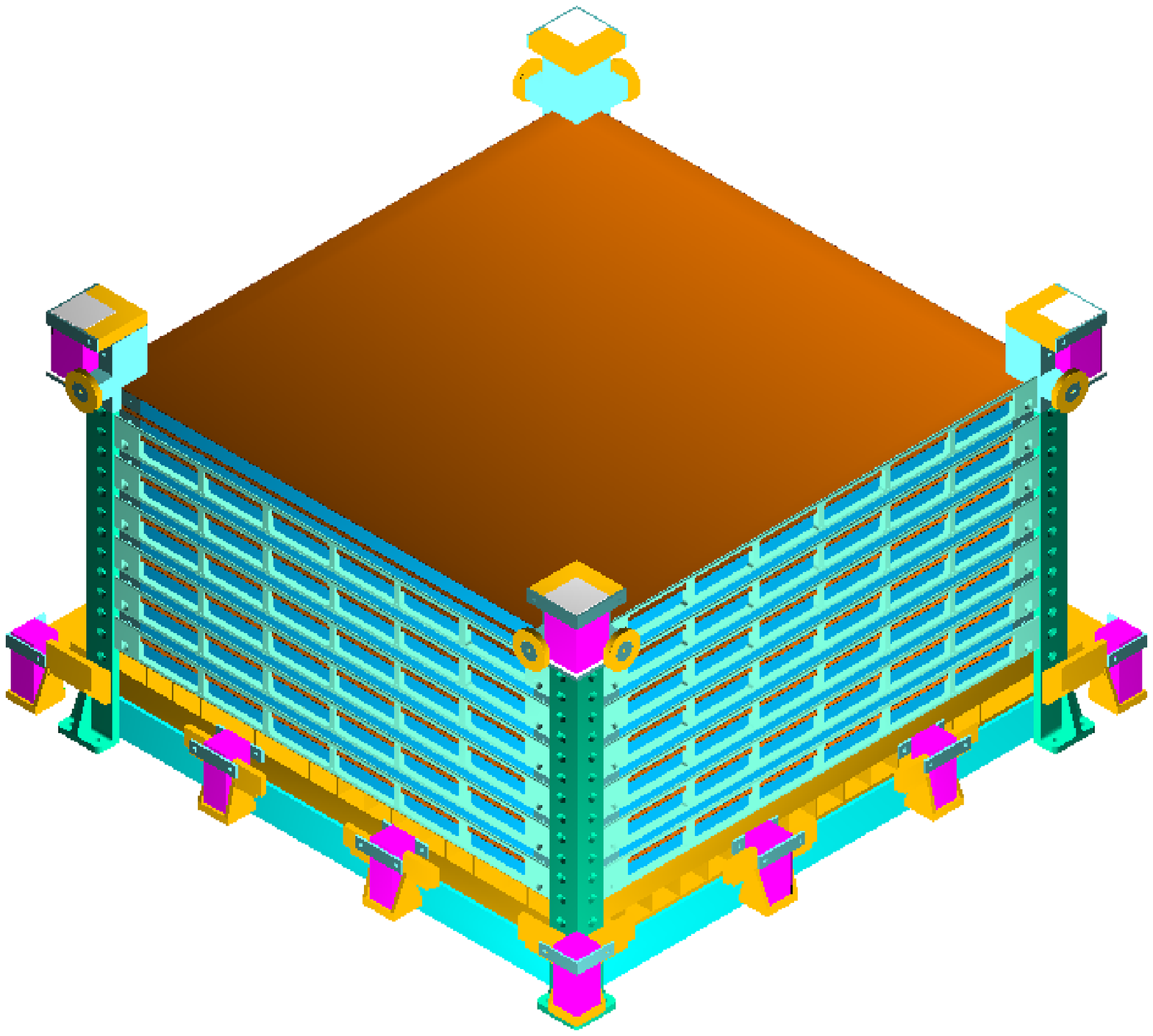,height=6.5cm,width=8.cm,angle=-90}}
\setbox222=\hbox{ \vspace*{-2.cm}
\psfig{figure=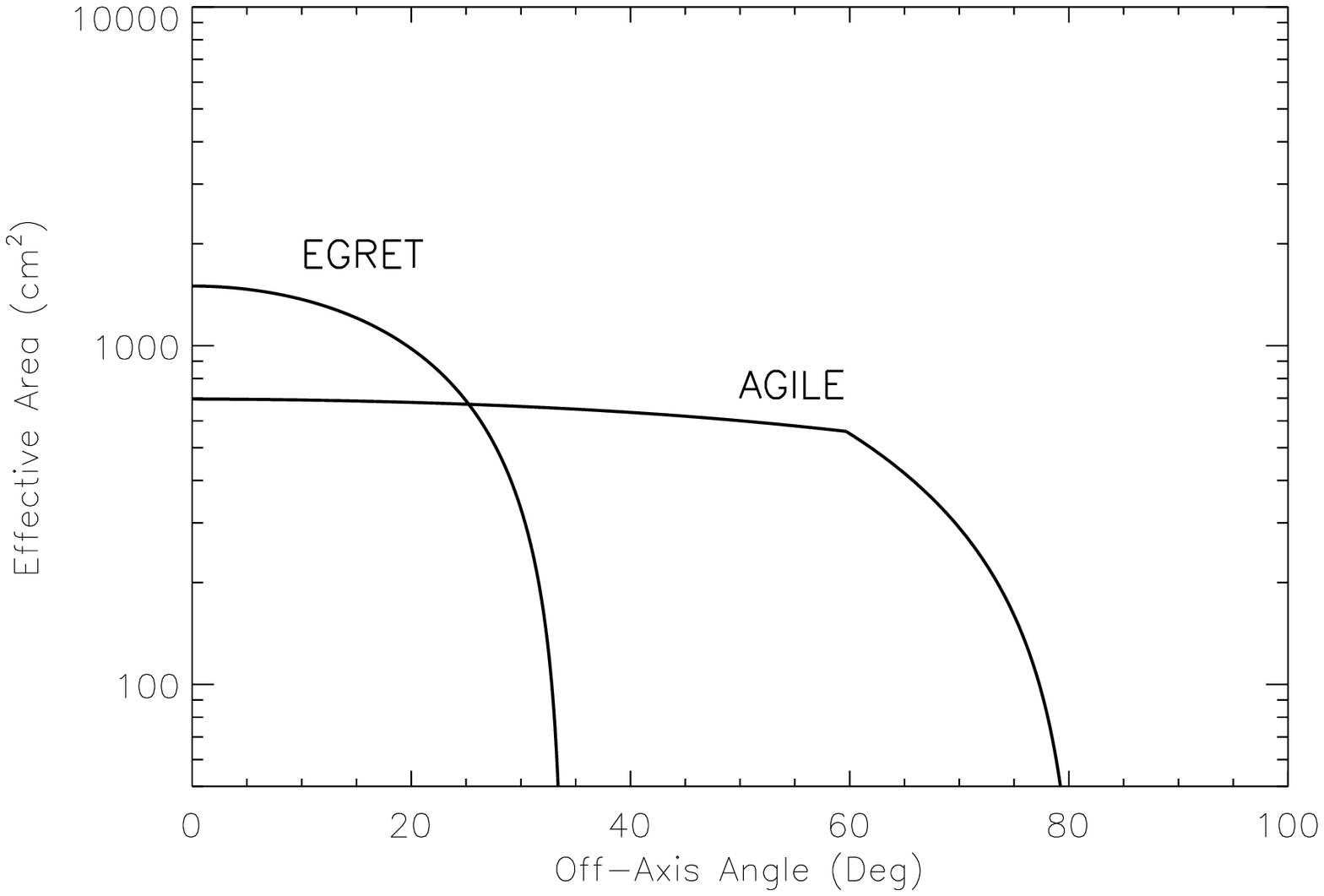,height=5.cm,width=6.cm}}
\centerline{\box111 \box222}
\caption{ {\it Left panel: }
Lateral view of the AGILE baseline instrument.
The anticoincidence system of plastic scintillator panels
(not shown)
 surrounds the detector made of 10 W-Si planes ($0.7 \, X_o$)
plus 2  more Si-only planes and a CsI mini-calorimeter ($1.5  \, X_o$).
The baseline payload size is $\sim 53 \times 53 \times 35 \rm \, cm^3$,
and $\sim 53 \times 53 \times 44 \rm \, cm^3$ for Super-AGILE. 
{\it Right panel: }
 Effective area at 1 GeV as a function of photon  incidence angle
for EGRET (Thompson et al., 1993) and AGILE. 
The segmented AC system  and AGILE's
trigger electronics allows detection of gamma-rays at large
incidence angles.
}
\end{figure}

\end{document}